\documentclass[11pt,twoside]{article}

\usepackage{asp2004}
\usepackage{epsf}
\usepackage{psfig}
\usepackage{lscape}

\begin{document}

\title{Astrometric Binaries in the Age of the Next Generation of 
       Large (Space) Telescopes}
\author{Rob P. Olling$^{1,2}$}
\affil{$^1$Dept. of the Navy, US Naval Observatory
3450 Massachusetts Ave NW, Washington DC, Washington  DC 20392-5420\\
$^2$Universities Space Research Association }

\begin{abstract} %%% Abstract to run on from here.
I analyze several catalogs of known visual and spectroscopic binaries
and conclude that a large number of binaries is missing in current
catalogs. Samples of the best studied (nearby and bright) stars
indicate that the true binary fraction may be as high as 95\%. A
preliminary analysis indicates that these binaries can affect the
astrometry significantly.
\end{abstract}

\section{Binarity: Past Present \& Future}

\citet[][hereafter referred to as DM1991]{DM1991} established the
multiplicity of G-type stars within about 22 pc. Their observational
dataset consists of thirteen years of radial velocity monitoring, in
combination with astrometric, visual and eclipsing binaries, for 164
G-type stars.

However, the pre-Hipparcos lists of wide binaries were rather
incomplete. Currently, the number of known binaries is about three
times larger. In fact, an analysis of the currently available binarity
and multiplicity data of nearby Hipparcos-selected G-type stars
indicates that as many as 95\% of primaries have a cataloged
companion. This binary fraction is almost twice as large as reported
by DM1991. In the remainder of this paper I will use the term binarity
loosely to mean binarity or multiplicity, where the companion can be a
star, a brown dwarf or a planet.

\subsection{Binarity: Present}
After the publications of the Hipparcos catalogs \citep{ESA1997},
several new sources of binarity have become available, which are
include in this paper.  Specifically, the following (compilation)
catalogs are used: HIP [the Hipparcos catalog \citep{ESA1997}], TY2
[The Tycho-2 catalog \citep{TYCH2}], TDS [the Tycho Double Star
catalog \citep{TDS}], SB9 [the 9$^{th}$ catalog of spectroscopic
binaries \citep{SB09}], EXOP [the catalog of confirmed extra-solar
planets \citep{exop}], INT4 [the 4$^{th}$ Catalog of Interferometric
Measurements of Binary Stars \citep{WI4}], and GCSN [the
Geneva-Copenhagen Solar Neighborhood Radial Velocity Survey
\citep{GCSN}]. This combined catalog contains 37,341 binaries, or more
than twice the number of Hipparcos multiples.  Angular separations are
available for 11,101 entries for the Hipparcos-only data, and for
almost three times as many (29,769) stars in the combined catalog. 

\begin{figure}[!ht]
   \vspace*{-5mm}
   \centerline{\resizebox{118mm}{!}{
      \plotone{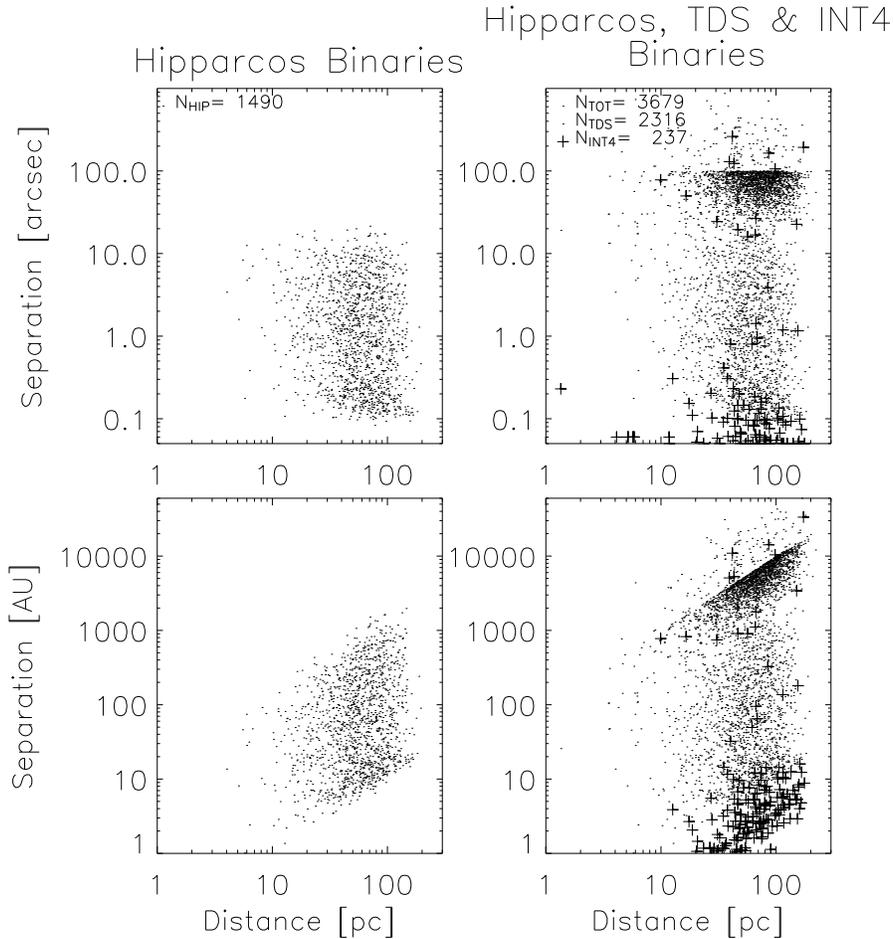}}}
   \vspace*{-5mm}
   \caption{ \label{fig:Hipp_T2DS_I4_Separations} Binary separations
   as recorded in the Hipparcos catalog (left column) and the combined
   catalog (right column). The top row presents the separations in
   arcsec, the bottom row in AU.  Many of the INT4 binaries (plotted
   as crosses) are still unresolved. Most of the objects in the INT4
   (crosses) at 0.1 arcsec separation are doubles at the Hipparcos
   resolution limit.The total number of binaries with a parallax
   accuracy better than 10\% and with a measured value for the
   separation are: 1491, 2316, 1292 and 5099, for Hipparcos-only
   binaries, TDS binaries, INT4 binaries and the sum of the three
   catalogs, respectively. }
\end{figure}

In figure~\ref{fig:Hipp_T2DS_I4_Separations}, I present, as a function
of distance, the binary separation in arcseconds (top) and AU
(bottom). The binaries included in the Hipparcos catalog [the combined
catalog] are plotted in the left-hand [right-hand] columns. In the
bottom row, the selection effect due to the distance of the stars are
apparent: at small distances, the large-separation binaries are
absent, while at large distances, the binaries with short semi-major
axes are missing.

Since the secondaries are drawn from the IMF (DM1991), one expects the
secondaries to be much fainter than the primaries, while faint
companions should outnumber the bright ones. However, since it is hard
to detect faint stars, the apparent binary frequency ($\beta$) might
be influenced by magnitude-based selection effects. Also, $\beta$ is
likely to drop with distance since companions of the same physical
separation are no longer resolved. Distant double stars are doubly
hard to detect: 1) because the are fainter, 2) because they are less
easily resolved.

\begin{figure}
   \vspace*{-5mm}
   \centerline{\resizebox{134mm}{!}{
      \plotone{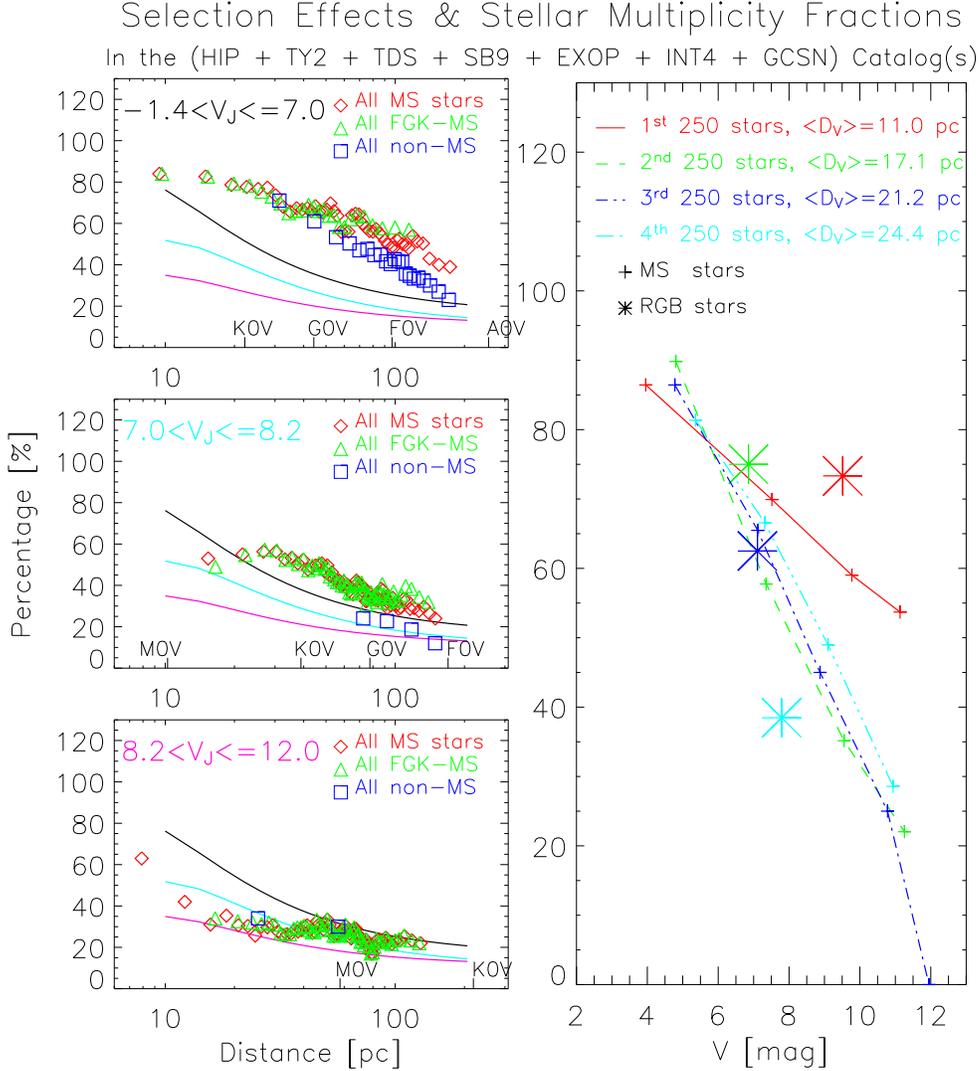}}}
   \vspace*{-5mm}
   \caption{ \label{fig:Hipp_T2DS_I4_GCSN_Multiplicity} Multiplicity
   fraction as a function of distance (left-hand panels) and apparent
   magnitude (right-hand panel) for the Hipparcos stars that have
   parallax errors less or equal than 10\%. The extrasolar planets
   contribute about 5\% to the nearest/brightest bins. The magnitude
   ranges for the left-hand panels are chosen to have an equal number
   of stars in each magnitude range. Plotted symbol correspond to a
   fixed number of stars. The averaged dependencies (as determined
   from the HIP+TY2+TDS sample) for the three magnitude intervals are
   plotted (drawn lines) in each of the left-hand panels. The
   main-sequence stars and the RGB stars behave in approximately the
   same manner.}
\end{figure}

Figure~\ref{fig:Hipp_T2DS_I4_GCSN_Multiplicity} shows that these
selection effects are clearly present in the data.  The left-hand
panels show $\beta$ decreasing rapidly with distance, and that the
brighter stars have higher average $\beta$'s, exactly as expected.
Also striking is the steep magnitude dependence of $\beta$ (right-hand
column).

I also created a G-star sample similar to the one of DM1991 by
selecting from the Hipparcos catalog all 204 GV stars closer than 22
pc that are cataloged as primaries. The behavior of $\beta$ (not
shown) for this sample is very similar to that of the full sample,
except that overall multiplicity level is increased, and that the peak
multiplicity reaches 95\%, while the minimum level is 65\%.

A proper analysis of this data in terms of the multiplicity of field
stars is beyond the scope of this paper. Even so, there is significant
evidence that $\beta$ is substantially larger than hitherto
determined. Because the most complete catalogs of binarity indicate
that $\beta$ is close to 100\% for the closest and brightest stars, I
surmise that the steep drop-offs towards larger distances and fainter
magnitudes are unphysical and in fact the result of selection
effects. Given that the sample most favorable for the detection of
companions has a multiplicity fraction of 95\%, I suggest that in fact
{\em all} stars are part of a multiple system. 

\section{Astrometry of Binary Systems}
Starting with a sample of well-measured Hipparcos stars and the period
distribution of DM1991, I generate a fake star catalog of the Solar
neighborhood. I estimate the masses of the systems based on the
observed color, and apply corrections based on the sample of measured
masses from the 6th Orbit Catalog. Positions along the orbits are
generated for these orbits, that roughly correspond to various planned
astrometric missions. The scatter incurred as a result of these
orbital motions are substantial and will contribute to the error
budget of a significant fraction of the stars.

%\acknowledgements %%% Text of acknowledgements runs on after this command.
%


\begin{thebibliography}{}
%
\bibitem[California/Carnegie Planet Search Compilation (2003)]{exop}
   California/Carnegie Planet Search Compilation.
   Extracted on 2003-09-09 from
   http://exoplanets.org/planet\_table.shtml
%
\bibitem[Duquennoy \& Mayor (1991)]{DM1991} 
   Duquennoy, A.~\& Mayor, M.\ 
   1991, \aap, 248, 485   (DM1991)
%
\bibitem[ESA (1997)]{ESA1997}
   ESA, 1997, The Hipparcos and Tycho Catalogues,
   ESA Publ. Division, ESTEC, the Netherlands
%
\bibitem[H{\o}g et al. (2000)]{TYCH2}
   H{\o}g E. et al., 2000, \aap, 357, 367
%
\bibitem[Fabricius et al. (2002)]{TDS} 
	Fabricius, C. et al., 002, \aap, 384, 180 
%
%\vspace{-0.5em}
%\bibitem[Hartkopf \& Mason (2004)]{Orb_6th} 
%   Hartkopf, W.~I. \& Mason, B.~D., 
%   \url{http://ad.usno.navy.mil/wds/orb6/orb6text.html\#stats}
%
\bibitem[Hartkopf et al. (2004)]{WI4} 
   Hartkopf, W.~I. et al., 
   http://ad.usno.navy.mil/wds/int4.html
%
\bibitem[Nordstr{\" o}m et al. (2004)]{GCSN} 
   Nordstr{\" o}m, B., et al., 2004, \aap, 418, 989
%
\bibitem[Pourbaix et al. (2004)]{SB09}
   Pourbaix, D., et al., 2004, \aap, 424, 727 
   http://sb9.astro.ulb.ac.be/intro.html\#Ack
%
\end{thebibliography}
\end{document}